# Light Concentrators for Borexino and CTF

L. Oberauer[a], C. Grieb[a,b], F. von Feilitzsch[a], I. Manno[c]

a) Technische Universität München, Fakultät für Physik, E15
   James-Franck-Str., D-85747 Garching

b) Virginia Polytechnic Institute & State University
   Physics Department, Robeson Hall
   Blacksburg, Virginia 24061

c) KFKI Research Institute for Particle and Nuclear Physics,
   Hungarian Academy of Sciences, PO Box 49
   H-1525 Budapest

**Abstract**

Light concentrators for the solar neutrino experiment Borexino and the Counting Test Facility (CTF) have been developed and constructed. They increase the light yield of these detectors by a factor of 2.5 and 8.8, respectively. Technical challenges like long term stability in various media, high reflectivity and radiopurity have been addressed and the concepts to overcome these difficulties will be described. Gamma spectroscopy measurements of the concentrators show an upper limit of 12 µBq/g for uranium and a value of 120 µBq/g for thorium. Upper limits on other possible contaminations like $^{26}$Al are presented. The impact of these results on the performance of Borexino and the CTF are discussed and it is shown that the design goals of both experiments are fulfilled.

**Motivation and requirements**

Borexino is an experiment dedicated to rare-event neutrino physics at low energies (i.e. in the sub-MeV range). The main goal of this experiment is real time solar neutrino spectroscopy via elastic neutrino electron scattering. The recoil electron spectrum of the mono-energetic (i.e. 861 keV) solar $^7$Be-neutrinos is of special interest. The Target as well as the detection medium are ~300t of a liquid scintillator. It consists of the solvent PC (1-2-4 Trimethylbenzene) with about 1.5 g/l PPO (2,5-Diphenyloxazole) added to it. The light yield of this scintillator is ~ $10^4$ photons for a beta-particle which deposits an energy of 1 MeV. The light is detected by photomultipliers (PMs) with their sensitivities matched closely to the maximum of the light emission spectrum which is at ~390nm. The chosen PPO concentration ensures large absorption lengths in the interesting wavelength region. A detailed description of the properties of the scintillator and the PMs is given in [1]. An overview about the construction of the Borexino detector as well as the background requirements is given in [2].

The highest possible light yield is crucial for Borexino. The signature of the $^7$Be recoil spectrum is a steep step-like function at the highest possible recoil energy of 660 keV. To make this step function visible against background events, the best possible energy resolution at this low energy is mandatory. In addition Borexino will apply position reconstruction of events by using information of the flight time of photons registered in the individual PMs. In this way a fiducial inner volume (of ~ 100t) can be defined and only events reconstructed inside this volume are counted as good neutrino candidates, whereas the by far dominant part of external background events should only create signals outside this volume. This external background may arise from gamma decays in the environment of the experiment (e.g. from the rock of the underground laboratory) or from the detector material (e.g. from the PMs) itself. In order to obtain a good rejection factor for such external background events a high light yield is necessary as the position resolution scales with the square-root of the number of photoelectrons measured. A third motivation to aim for an optimal light yield comes from the background contribution which may be expected from alpha-decays occurring in the scintillator liquid. They stem from α-decays present in the radioactive U- and Th-chains. As the alpha events are quenched in a liquid scintillator by a factor ~10, most of the multi-MeV signals fall inside the region of interest for solar $^7$Be neutrino detection, which is below ~800 keV [2]. Identification of alpha-events is possible via the analysis of the pulse shape of the individual events. Like energy and position resolution, pulse shape analysis also depends significantly on the strength of the signal, i.e. on the number of photoelectrons.

These considerations show that an optical coverage as high as possible is desired. However, the total number of PMs used is limited by the fact that they contribute to the external background themselves. Therefore it was decided to equip the PMs with light concentrators which should allow a high light yield by keeping the external background low enough to allow solar neutrino measurements. The same considerations apply also the CTF, where scintillator properties can be studied with high precision [3]. The geometry and the environmental medium in the CTF differ completely from Borexino. In the CTF, only 100 PMs collect the light from a scintillator sphere with a radius of ~1m. In Borexino, the concentrators are immersed in PC, an organic solvent, but in the CTF the concentrators are immersed in deionized water. Hence the requirements on the materials as well as the geometric form differfor both cases.

The main specifications for the concentrators in Borexino are: highly efficient and uniform light collection from the scintillator region, low background contribution (compared to other detector materials) in the energy spectrum of solar neutrinos, and long term stability against corrosion in either PC or deionized water as environmental medium.

**Shape of the concentrators optimized for Borexino and CTF**

The profile of the light guides for Borexino and CTF is designed with the "string method" (e.g. [4]). These concentrators collect the light from the scintillation vessel uniformly. The radius of the scintillator vessel in Borexino is 4.25 m and the front of the PMs are 6.52 m from the center of the detector. Light emitted from the scintillator enters a concentrator over a range of angles from zero degrees to some maximum angle $\delta_{max}$. All photons entering with $\delta < \delta_{max}$ are directed, after one reflection at most, to the rim of the exit aperture (Fig. 1). Photons from other directions should therefore not be accepted by the concentrator, a phenomenon that is intuitively appealing and works perfectly in two dimensions and nearly perfectly in three dimensions. The surface of the light guide is designed in a way that light emitted from a spherical effective volume (r = 4.45 m) is uniformly collected on the light guide's exit aperture, as opposed to a parabolic mirror which would focus incoming light on a focal point. In this way the scintillation light is distributed on the curved photocathode of the Thorn EMI 9351 PM, which is shaped like a spherical section (the maximum radius of the cathode is 9.5 cm and its curvature radius is 11.0 cm, the height of the spherical section is equal to 5.5 cm). The larger effective radius of 4.45 m was chosen to allow for small misalignments of both the PMs as well as the scintillator vessel. In two dimensions both the photocathode and the effective fiducial volume are circular. In the "string method" the light ray, which is tangential to the effective volume, is directed into the extreme point of the exit aperture after one reflection on the light guide profile or it is reflected in such a way that it is tangential to the photocathode also (Fig. 2). One may design the profile using a string whose length is constant (Fig. 2). The length of the string is determined by fixing one end of the string to an extreme point of the spherical effective fiducial volume and the other end is fixed to an extreme point of the photocathode. The length of the string is chosen so that it is taut when it is pulled to the other extreme point of the photocathode. The whole string consists of two arches and one straight section. This is similar to the so called gardener's method of drawing an ellipse, where the profile is drawn keeping the string taut.

The calculated transmission curve of the light guide is shown in Fig. 3 for three dimensions. It shows the collection efficiency of the concentrator as a function of the angle of incidence $\delta$. The radii of the entry and exit apertures are 15.1cm and 9.5cm respectively. The critical angle of incidence is $\sim 44^\circ$ (Fig. 3). For this calculation an average reflectivity of 86% of the cone (i.e. the polished aluminum surface) was assumed. With this value the collection efficiency is $\sim 0.88$ within a wide range of incidence angles $\delta$ and coincides with the experimentally observed value (see discussion below). The ideal shape of the concentrator implies that the aperture reduces slightly again towards its opening. We truncated the cones where the diameter is maximal, reducing the length of the light guide by 1.5 cm. The loss in light yield is 2.25 % only. This decision was based on cost considerations. Sticking to the ideal form would have meant a doubling of the cost for the spinning process. The distance between the center of the scintillator vessel and the PM (the plane which contains the maximum radius of the cathode) is 6.25 m. The geometric amplification factor is defined as the ratio of solid angle coverage of the bare PM and a PM with mounted light guide. One may calculate this geometric amplification factor with the above mentioned numbers to be 2.7 for the Borexino light guide. Figure 4 shows the average number of reflections for photons as a function of the angle of incidence, calculated by a raytracing program.

The CTF concentrators are shaped using the same method, yet the different geometry of the CTF produces a different result. In the CTF the distance from the PMs to the center of the detector is 3.3 m, the effective radius chosen is 1.2 m (the real radius of the scintillator sphere is ~1 m). Therefore the critical angle of incidence of the CTF is $27^o$ only and the shape of the CTF-cones is more oblong compared to the Borexino type. With a length of 57 cm and a radius of 25,3 cm the concentrators provide a geometrical amplification factor in the CTF of ~10 for a single PM. This makes it possible to study the properties of liquid scintillators with large masses of ~4 to 5 tons at very low energy thresholds of about 20 keV (see discussion below) in the CTF with the use of only 100 PMs at a large distance (to reduce background) of 3.3 m.

Borexino uses 2214 PMs, 1870 of which are equipped with concentrators. Hence, the total effective geometrical coverage is 29.8%. The motivation for using 400 PMs without concentrators is to improve the separation power between point like events (e.g. due to neutrinos) and the signal created by muon-tracks in the buffer region [2], which can produce a non-negligible background contribution as observed in CTF-runs [5]. High energy muons traversing the buffer region create Cherenkov as well as scintillation photons (the latter is reduced due to the fact that there is no PPO present in the buffer liquid) along their tracks. Many of these photons will be not accepted by the light concentrators as the angle of incidence is too large. However, PMs without cones can observe those photons. Therefore the fraction of photoelectrons seen by PMs with cones divided by the number observed in PMs without cones is significantly lower for muon events which have tracks through the buffer region as compared to point like events emerging from the scintillator region. The number of PMs with and without concentrators have been optimized in Monte-Carlo calculations.

**Construction and reflectivity of the concentrators for Borexino and the CTF**

The reflectivity of the mirror's surface is crucial for the application. The wavelength regime important for the detection of the scintillation photons in Borexino and CTF is between about 370nm to 450nm. This window is determined by the emission spectrum of the PPO-molecules and the efficiency function of the PMs. Within this wavelength band aluminum and silver are chosen to act as reflective layers. Both have a bulk reflectivity of ~90% at around 400nm. Any other metal shows a significantly decreased reflectivity at this wavelength.

As de-ionized, high purity water may corrode aluminum quite efficiently silver was used as a reflective layer for the CTF-concentrators. On the other hand silver reacts with PC and the surface gets a purple color. For this reason aluminum was used for Borexino. During the development of Borexino a method to produce aluminum cones which can also resist ultra-pure water for ~10 years (see text below) was found. This development was important; it allows the installed PMs and cones to be cleand with highly purified water after installation. In addition, Borexino may be filled with water which later can be replaced by PC. This may have certain strategic advantages.

*Construction of CTF concentrators:* The bulk material chosen are UV-transparent acrylic plates with polished surfaces. These were tested for radioactive impurities by gamma-spectroscopy with Ge-detectors. Only upper limits on U and Th in the range of $10^{-9}$ g/g were found. The measured values for natural K of ~$10^{-7}$ g/g are acceptable for the purpose of the CTF as the contribution of background events due to $^{40}$K gamma rays with 1.46 MeV at this level is negligible. Due to the enormous passive shielding (water in the CTF, pure PC in Borexino) the highly penetrating 2.6 MeV gamma emission from $^{208}$Tl is the most dangerous source of external background. As $^{208}$Tl occurs in the lower part of the thorium-chain the Th-concentration is the most crucial parameter. After cutting, the plates were heated in an oven to release humidity. Both surfaces have been covered with a removable plastic layer for protection. The plates were shaped by thermal deep drawing against a positive form which has the shape calculated by the string method described in the chapter before. During this process only the inner layer of the plastic was removed. The outer layer still shielded the surface against the exposure of radon daughters which are attached to air-molecules in form of aerosols and dust. After the drawing process the cones were cut on both ends. Four acrylic holders have been glued to each concentrator at the rim for linking the cone to the PM. All cones were carefully sealed in plastic bags. Slightly before the evaporation of silver onto the outer surface the plastic layers were removed. Then a thin layer of copper was evaporated onto the silver. Several laboratory tests found that this double-layer shows very good properties concerning adhesion on acrylic plates and a strong resistance against corrosion in de-ionized water. Finally, a thin acrylic layer was sprayed on the outer surface for additional protection. The cost for one CTF-concentrator was ~230€ which is only a small fraction of the cost for one encapsulated PM plus electronics.

*Long term stability of CTF concentrators:*

In accelerated aging tests in the laboratory, the long term stability of the material composition of the cones has been tested. These tests were performed using de-ionized water with the same quality used in the CTF, but at higher temperatures. According to general laws of corrosion it is assumed that an increase of 10 degrees corresponds to a time factor of ~2. During the development of the Borexino cones this relation was tested by measuring the corrosion current as a function of temperature (see below in the Borexino cone section) and it was found to be valid in good approximation. The CTF concentrators were installed during July 1994. They were used from December 1994 until March 1997 and no deterioration in the light yield in PMs which survived this period was observed. After emptying the CTF tank, several concentrators were replaced by new ones, because they had developed a thin gray layer on the outside. This had no impact on the light yield, since it was on the outside of the concentrator. Nevertheless, those were exchanged with new cones that have an additional silver layer outside instead of the acrylic spray. Some of the used cones had scratches on the surface, presumably caused during the installation period. The concentrators were then used in water from 1999 until March 2001. Deterioration of the light yield was not observed yet again. A new scintillator (PXE) was tested in that CTF campaign. During the draining of the CTF, the scintillator sphere (a thin nylon balloon) was accidentally destroyed. Many liters of PXE (an organic solvent) went into the CTF water and came in contact with most of the concentrators which were placed on the lower hemisphere. Due to this contact basically all of these cones developed several cracks within a few days. This effect was confirmed in laboratory tests. Many of the CTF cones could be glued back together. Some had to be replaced by new concentrators. Since fall 2001 the CTF is running again and no deterioration of the concentrators and hence the light yield has been observed.

*Construction of Borexino concentrators and long term stability:* Due to the obvious incompatibility of acrylic glass with organic solvents, the Borexino type concentrators were made from aluminum. The most difficult problem was to find aluminum pure enough concerning radioactive elements (especially in thorium), relatively inexpensive and easy to machine. After finding the right bulk material (see text below), smooth thin 3mm plates were produced. Round discs were cut from the raw plates. Those discs were then shaped in a spinning process. To improve the smoothness of the surface, the mirrors were ground and polished on the inside. Afterwards they were anodized in several steps to a deepness of ~7μm. This makes them extremely inert without reducing the reflectivity too much. Accelerated aging tests showed that those concentrators have long life times even in deionized water. One cone had been immersed in hot water for several months (corresponding to a life time of ~10 years at $15^0$C) and only showed corrosion at the holding points where the anodizing layer is compromised. The reflectivity has been measured after the corrosion test and no decrease in the light yield within the quoted uncertainties has been found. In a corrosion experiment, in which the material to be tested was used as electrode, the corrosion current was measured as a function of the temperature. The corrosion current allowed to determine the speed at which the oxidization process takes place. This experiment yielded a corrosion rate which is equivalent to a loss of material of $(3.0 \pm 0.5)$ nm per year. In addition we tested the reflectivity before and after the corrosion tests, and found no decrease within the error bars. The material, immersed in PC, showed no visual deterioration at all for several years of testing. This was tested by visual observation and by weight measurements. The cost for one Borexino concentrator was ~100€, whereas the total cost for one PM-channel (encapsulated PM plus electronics) amounts to ~ 2000€. As the amplification factor is about 2.5 (see discussion below) the use of light concentrators to reach a high light yield is fully justified; in order to obtain the same light yield without concentrators we would have needed an additional ~10M€.

*Reflectivity of the CTF and Borexino concentrators:*

The cone diameters as a function of the height have been found to be in agreement with the theoretical curve within ~1mm. This minor deviation from the design specification introduces only a negligible loss in the collection efficiencies. The Borexino concentrators show a less smooth inner surface compared to the acrylic CTF-type. A parallel laser beam hitting the surface under $45^o$ shows typically a broadening of ~$3^o$ after one reflection. According to calculations and in agreement with our measurements (see below) this has only a marginal influence to the photon collection efficiency.

The photon collection efficiency of the mirrors for the CTF and Borexino concentrators were measured with scintillation photons. A small (5cm in diameter) glass sphere, filled with a PC/PPO solution at the same concentration as in Borexino was used. Gamma rays with 660 keV emitted from a $^{137}$Cs gamma-ray source hit the scintillator sphere. Compton-scattered photons with $\approx 180°$ scattering angle are detected in coincidence by a NaI detector placed behind the source and a PM close to the sphere. At several distances (varying from 29cm to 98cm) a second PM is placed. The ratio of triple coincidences $n_3$ between all three detectors and double coincidences $n_2$ between NaI and near PM was measured. The comparison of data with and without concentrator in front of the far detector at different distances between source and PM allows the determination of the overall collection efficiency. The average number $\mu_0$ of photoelectrons registered in the case of no cone placed in front of the PM is

$$\mu_0 = const \frac{\pi r^2}{4\pi d^2},$$

where $r$ is the radius of the PM entrance window ($r=9.5$cm), $d$ is the distance between source and PM, and the constant includes the light yield of the scintillator events in the energy window selected by the double coincidence and the efficiency of the PM. The corresponding number $\mu_+$ in the case of a concentrator in front of the PM is $\mu_+ = const \frac{\pi R^2 \varepsilon}{4\pi (d-l)^2}$, where $R$ and $l$ are the radius of the entrance side ($R=15.1$cm) and the length of the concentrator respectively, and $\varepsilon$ is the collection efficiency to be found. The average number of photoelectrons registered can be determined experimentally by measuring the ratio $n_3/n_2$ and by the relation $\mu = -\ln(1 - \frac{n_3}{n_2})$, which is derived from Poisson's statistic. The threshold of the PM was adjusted to be fully sensitive to single photon pulses. The distance $d$ was chosen in such a way, that the values of $n_3/n_2$ vary between 0.3 and 0.6 in order to avoid large uncertainties. The comparison of measured values of $\mu_0$ and $\mu_+$ directly delivers the quantities *const* and *const* $\varepsilon$. Assuming that the value of *const* is the same in the measurement with and without cone one gets the collection efficiency $\varepsilon$. The result for the Borexino type concentrators is $\varepsilon = 0.88 \pm 0.04$ and by coincidence the same result was obtained for the longer CTF-cone. The direct measurements of the reflectivity of anodized aluminum done by the manufacturing company showed values around 80%. This reduction compared to the theoretical maximal value of 91% (at 400nm) is due to the anodizing process. As the average angle of incidence in our measurements (and also in the case of CTF and Borexino) is significantly lower the effective reflectivity should be increased and our experimental results are plausible. Monte-Carlo calculations agree with this experimental result if the average reflectivity chosen is $r = 0.86$ (Fig. 3). The laboratory measurements were performed in air. The average angle of incidence of light on the glass of the PM is increased when a concentrator is placed in front of it. This may reduce slightly $\varepsilon$ as the reflectivity of the glass increases. In CTF and Borexino the medium is water and PC respectively and the indices of refraction between medium and glass match much better. Hence in the real application $\varepsilon$ may be slightly higher as measured in the laboratory. Neglecting this effect and using the measured laboratory values the amplification factors in Borexino and CTF are 2.5 and 8.8, respectively. By 2001 basically all concentrators were mounted in the Borexino detector. The first test measurements using a small scintillator sphere loaded with Radon showed good results concerning the light yield. In the CTF the photoelectron yield is ~300pe/MeV [3]. The threshold is set for a coincidence of 6 PMs. This threshold is high enough to reduce accidental trigger events due to the thermal noise of the PMs to < 1 count/day. As this threshold is equivalent to ~6pe, the achieved energy threshold in the CTF is ~20keV.

**Radiopurity of the concentrators**

Borexino is an ultra-low background detector. It will observe 12 to 55 neutrino events per day (depending on the neutrino oscillation scenario assumed) mainly due to $^7$Be-neutrino electron scattering in a fiducial mass of 100 tons of scintillator in the recoil energy window between 250 keV and 800 keV [2]. In the energy window 0.8 MeV to 1.5 MeV solar neutrinos from the pep-reaction and the CNO-cycle should be counted with a rate between 0.7 and 3.5 events per day. To distinguish the neutrino signal from the radioactive background, all components of the detector are carefully selected to avoid radioactive contamination. Aluminum usually has a small content of thorium (and its daughter isotopes), which can be removed at great cost. The contents of uranium in aluminum are typically small. The raw aluminum was selected by making field measurements of different types of aluminum. Unfortunately, several batches of aluminum had to be used for the production. The measurements were done in the underground laboratory in Garching by gamma-spectroscopy with a 115 cm$^3$ Germanium detector and in Heidelberg at the MPIK. It was found that the equilibrium in the uranium as well as in the thorium decay chain is usually broken, presumably during the production process. The concentrations of radioactive isotopes vary between the different types of aluminum. The content of $^{208}$Tl is most important to the experiment, because its decay produces a 2615 keV gamma-ray with a high branching ratio. Gamma rays with such high energies can penetrate the shielding of Borexino. The concentrations of thorium in our aluminum samples varied between 7 ng/g ($\hat{=}$ 28µBq/g) and 220 ng/g ($\hat{=}$ 890µBq/g). Also samples of 'Kryal' (99.9995% purity) have been analyzed. Their average value of ~ 30 ng/g ($\hat{=}$ 120 µBq/g) in thorium with only a small variability was quite good. However, the rather high costs of the bulk material and severe technical difficulties to process the soft material to plates which can endure the further processes (i.e. spinning, polishing, and hard anodization) caused us to stay with the usual 99.8% pure aluminum. The average value for the material finally used was ($\hat{=}$ 730µBq/g). Only upper limits on the concentrations in uranium (< 1ng/g or 12 µBq/g) below $^{226}$Ra and potassium ( < 10 µg/g or 320 µBq/g) could be found. In principle, $^{26}$Al, which is produced in reactions of cosmic rays in the atmosphere, could play a role due to the 1809 keV gamma emission. However, the measured upper limit of $< 4 \cdot 10^{-13} \, g/g$ ($\triangleq$ 27 µBq/g) for $^{26}$Al in our samples excludes this possibility. The typical values for uranium above $^{226}$Ra were around 100 ng/g ($\triangleq$ 1.2 mBq/g), clearly showing the break of the radioactive equilibrium (radium is depleted). Fortunately, the high energy lines of the uranium chain are below $^{226}$Ra and hence no contributions from this chain has to be expected for the external background. The same holds for the $^{235}$U chain. Here only an upper limit of 2 ng/g ($\triangleq$ 130 µBq/g) could be determined in gamma spectroscopy. The energy of the $^{40}$K-line with 1460 keV is too low to penetrate the shielding of Borexino in a substantial way. Hence, even though the limit of 10 µg/g is rather high, no contribution to the background has to be expected. Therefore all background events in the energy region of interest are basically due to the 2.6 MeV $^{208}$Tl-line.

The specification for the choice of material was to restrict the contribution of background events in the $^7$Be-neutrino window to an almost negligible level. This was studied in Monte-Carlo calculations. The external background in the fiducial volume (i.e. in the inner 300t of scintillator) due to the aluminum used in the light concentrators was determined to be $n_{7Be,conc} = (0.08 \pm 0.04)/day$ in the energy interval 0.25 - 0.8 MeV ("$^7$Be-neutrino window"), see Table 1. This corresponds to ~ 0.2 % of the solar neutrino signal in this energy region for no-oscillations and ~ 0.3 % for the LMA-solution. This contribution is lower than the unavoidable in-situ cosmogenic background due to the production of radioactive $^7$Be-nuclei (~0.35/*day*) [6] which occurs in spallation processes with high energy muons intersecting the scintillator sphere. Hence the background specifications were fulfilled. This contribution in this window is a little lower compared to the background introduced by the PMs itself ( $(0.15 \pm 0.05)/day$ ).

With the amplification factor being ~2.5 the use of the concentrators is justified also from this point of view: in order to achieve the same light yield without cones by additional PMs the total external background would increase to ~ 0.4 /$day$.

The background due to the concentrators in the so-called pep-neutrino window (0.8 - 1.5 MeV) is $n_{pep,conc} = (0.20 \pm 0.05)/day$ and accounts only for ~ 25 % of the total background PM plus cone. Since the pep-neutrino flux is much lower, the background to signal ratio here is ~5.7% for no-oscillations and ~11% for the LMA-solution. With that, the radiopurity of the concentrators is good enough to allow solar neutrino measurements also in this energy region. Although it should be expected that contributions from other sources (like the PMs) to the total external background will exceed this number. In addition the in-situ cosmogenic background due to $^{11}$C-decays will dominate with ~ 11/$day$ [Tanja]. Also the requirements to the internal radiopurity of the scintillator are very challenging. In order to allow pep-neutrino measurements the concentrations in uranium and thorium should be below $10^{-17}$ g/g or $10^{-13}$ Bq/g. Therefore the feasibility of a solar pep-neutrino measurement is not clear yet.

**Conclusions**

In the context of the solar neutrino experiment Borexino and the Counting Test Facility CTF at the Gran Sasso underground laboratory light concentrators with high photon collection efficiencies have been realized. For Borexino and the CTF low energy thresholds, a good energy and position resolution as well as a sufficient beta-alpha separation is essential. The developed concentrators allow us to achieve these goals with a relatively small number of PMs. The amplification factors for the CTF and Borexino are 8.8 and 2.5, respectively. The costs are only about 12% (CTF-concentrator) and 5% (Borexino-concentrator) compared to the cost of a fully equipped PM. Hence the CTF-and Borexino-detectors performances are substantially enhanced at low costs. No appreciable additional external background to the experiments is introduced. The devices are compatible withthe organic solvent PC as well as with highly purified water for a long (~10 years) period. Fig. 5 shows a CTF light concentrator and Fig. 6 shows a Borexino light concentrator, both mounted on a PM.

**Acknowledgement**

This work was supported by the "Bundesministerium für Bildung, Wissenschaft, Forschung und Technologie" (BMBF), the "Deutsche Forschungsgemeinschaft" (DFG), and the "Technischen Universität München" (TUM). We acknowledge the technical support from the companies "Klobe&Sohn" (evaporation techniques), "Metalldrückerei Wender" (spinning processes), and "Süd-Eloxal" (hard anodization of aluminum) . Our special thanks to our former colleague Dipl.-Ing. K.H. Schuhbeck. We would like to thank the whole Borexino collaboration, especially L. Cadonati and J. Maneira for Monte-Carlo calculations, G. Heusser for gamma spectroscopy measurements performed in Heidelberg, and the crews which did a marvelous job installing the PMs and concentrators in Borexino and CTF.

**Table and Figure captions**

Table 1: Background in Borexino: Comparison between the contributions of the lightguides and the PMs. The $^7$Be-ν window will be 250 – 800 keV and the pep-ν window 800 – 1300 keV.

Fig. 1: String Cone for Borexino. The entry and exit apertures' radii are 16 cm and 9.5 cm respectively. The length of the light guide is 27 cm. Due to practical reasons the manufactured concentrators are truncated at a height of 23 cm. The critical angle of incidence is $\delta_{max} = 44^o$. For values above $\delta_{max}$ the light will be reflected. The geometrical amplification factor of the concentrator in the configuration of Borexino is 2.7.

Fig.2: Illustration of the string method for constructing the shape of the concentrator (see Text). A light ray originating at the boundary of the region of interest is reflected at point p ant hits the PM's photocathode tangentially. This example corresponds to the case of the CTF. Here the geometrical amplification factor (coverage ratio) is ~10.

Fig. 3: The calculated transmission curve (i.e. the collection efficiency for photons as a function of the angle of incidence) of the string cone designed for Borexino. The critical angle of acceptance is at ~44$^o$. Data points are obtained in a Monte-Carlo calculation. The reflectivity $r$ (for vertical direction of incidence) was handled as a parameter. For $r = 0.86$ the data points coincide with the experimental value of 0.88 for the overall collection efficiency.

Fig. 4: Calculated number of average reflections of a photon inside the Borexino-concentrator until it hits the photocathode of the PM as a function of the angle of incidence. Obviously the ideal case of a two-dimensional cone with one reflection only is achieved within good approximation.

Fig. 5: CTF light concentrator mounted on PM.

Fig. 6: Borexino light concentrator mounted on PM.Tables and Figures

|             | $^7$Be-ν window counts/day | pep-ν window counts/day |
|-------------|---------------------------|-------------------------|
| Lightguides | 0.08+-0.04                | 0.20+-0.05              |
| PMs         | 0.15+-0.04                | 0.54+-0.08              |
| Expected rate (LMA) | 33                | 0.7 – 3.5               |

Table 1: Background in Borexino: Comparison between the contributions of the lightguides and the PMs. The $^7$Be-ν window will be 250 – 800 keV and the pep-ν window 800 – 1300 keV.

# Borexino Light Guide

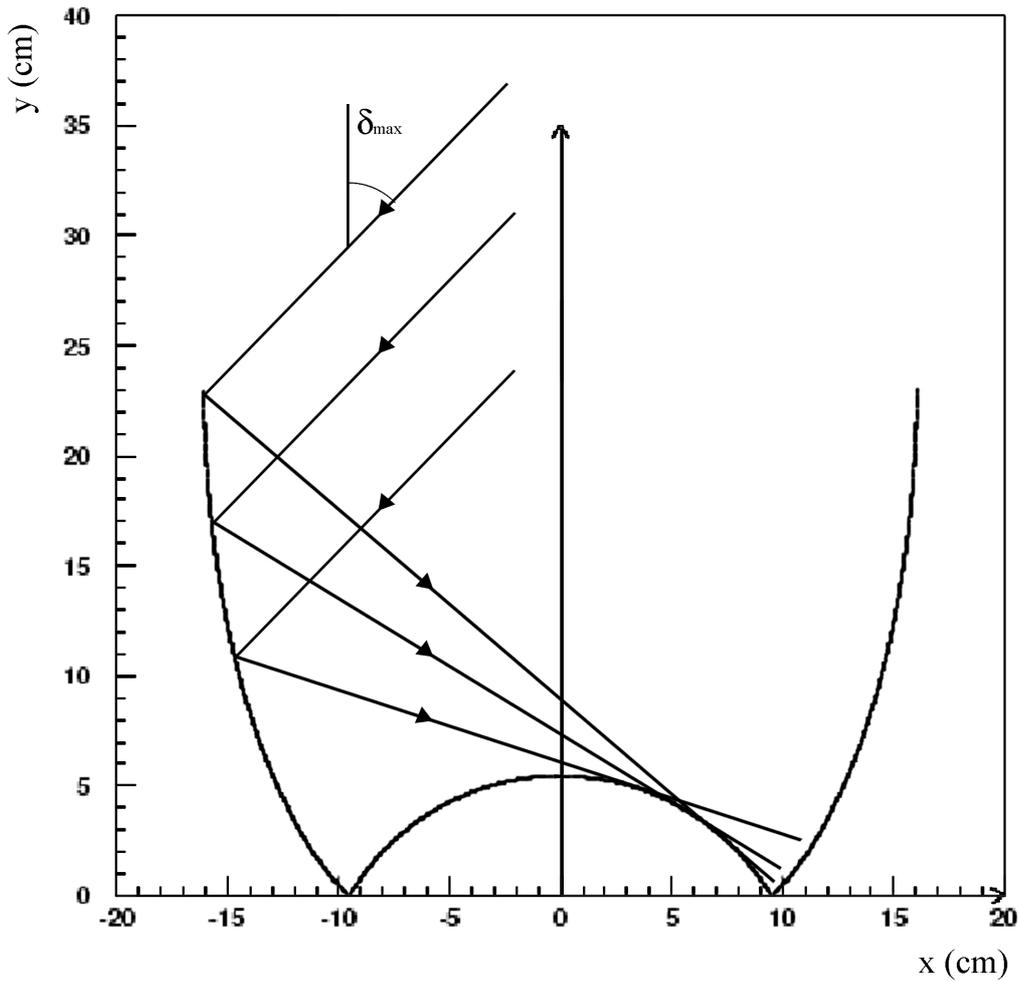

Fig. 1: String Cone for Borexino. The entry and exit apertures' radii are 16 cm and 9.5 cm respectively. The length of the light guide is 27 cm. Due to practical reasons the manufactured concentrators are truncated at a height of 23 cm. The critical angle of incidence is $\delta_{max} = 44°$. For values above $\delta_{max}$ the light will be reflected. The geometrical amplification factor of the concentrator in the configuration of Borexino is 2.7.

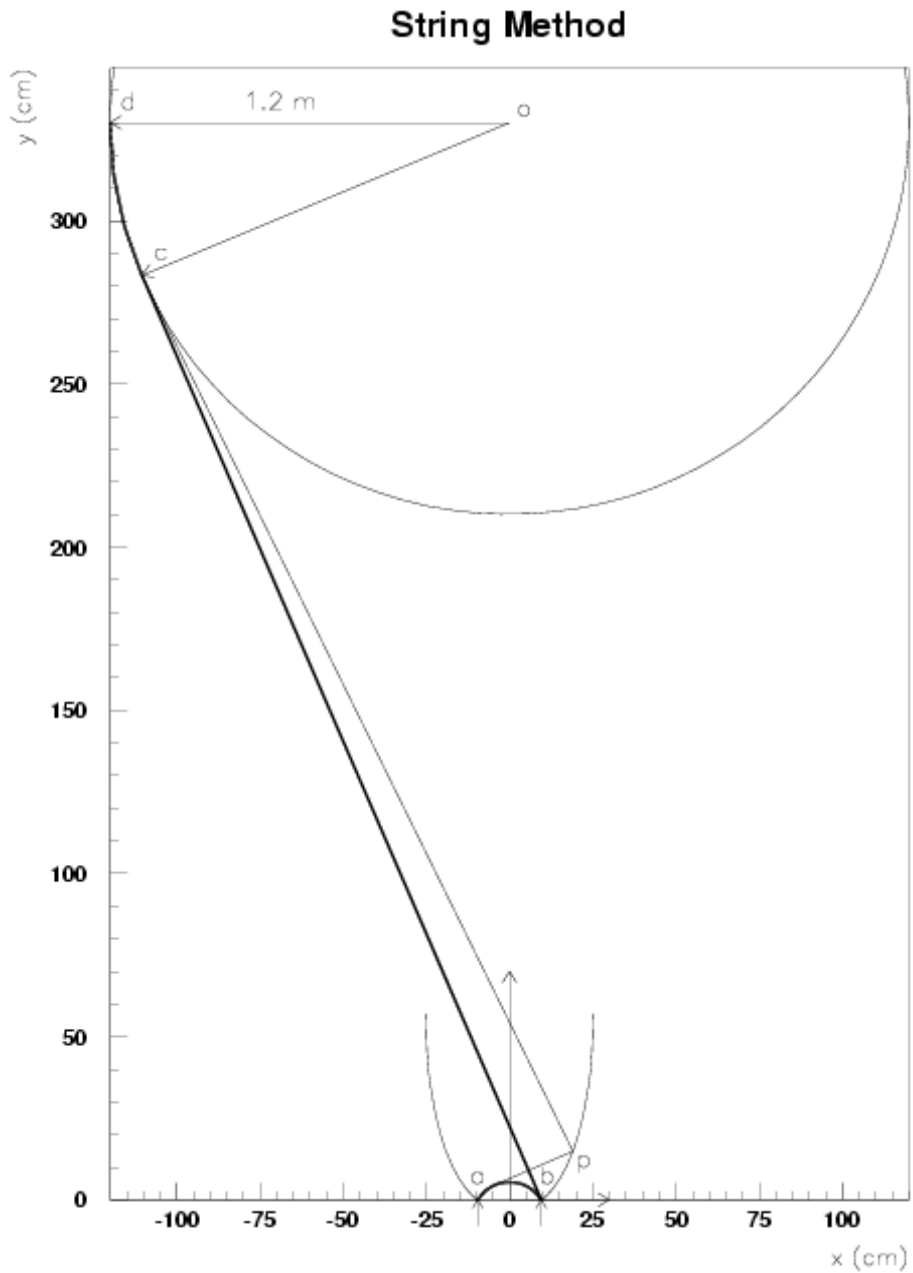

Fig.2: Illustration of the string method for constructing the shape of the concentrator (see Text). A light ray originating at the boundary of the region of interest is reflected at point p ant hits the PM's photocathode tangentially. This example corresponds to the case of the CTF. Here the geometrical amplification factor (coverage ratio) is ~10.

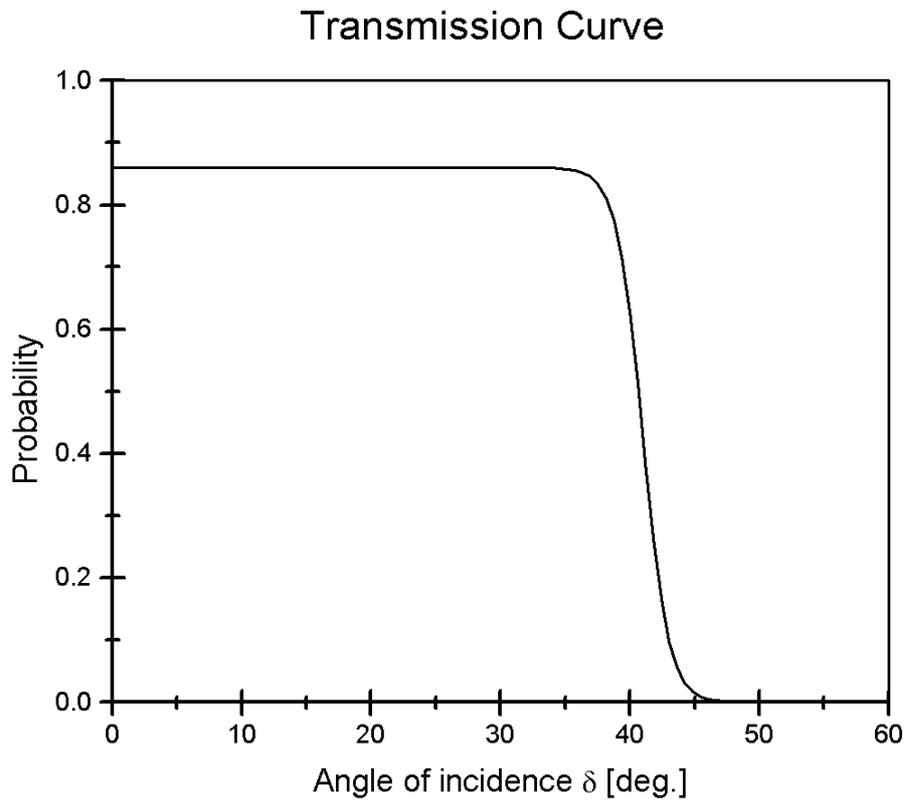

Fig. 3: The calculated transmission curve (i.e. the collection efficiency for photons as a function of the angle of incidence) of the string cone designed for Borexino. The critical angle of acceptance is at ~44°. Data points are obtained in a Monte-Carlo calculation. The reflectivity $r$ (for vertical direction of incidence) was handled as a parameter. For $r = 0.86$ the data points coincide with the experimental value of 0.88 for the overall collection efficiency.

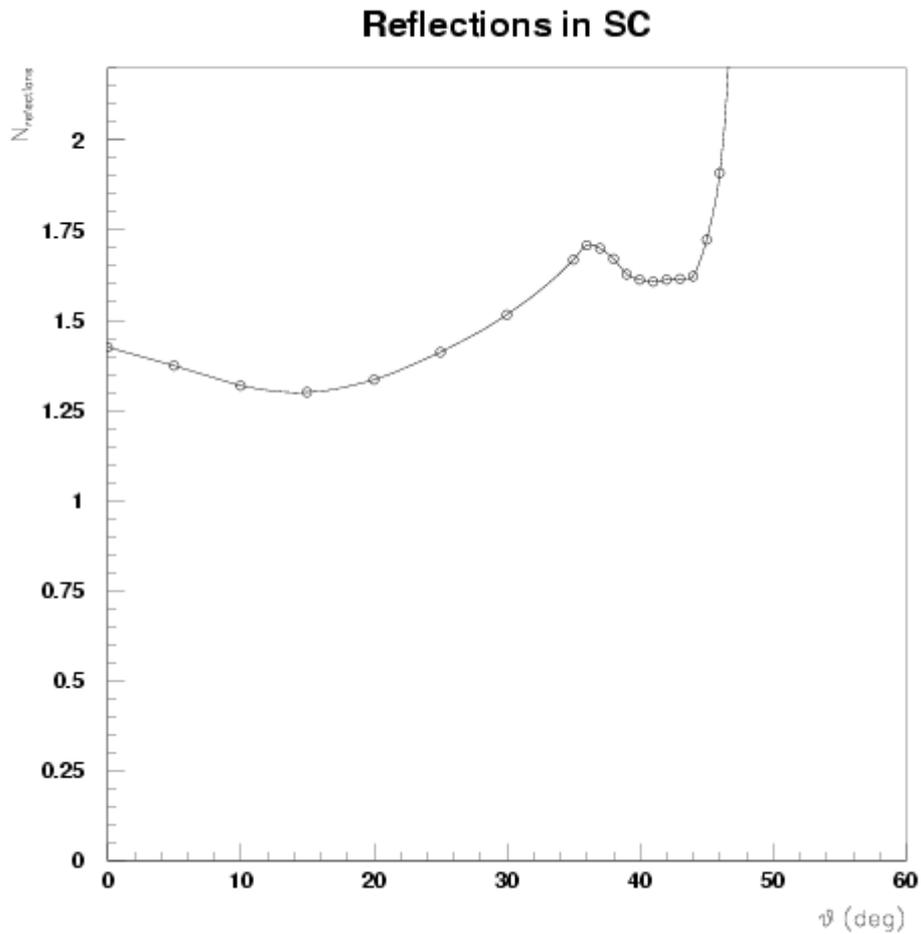

Fig. 4: Calculated number of average reflections of a photon inside the Borexino-concentrator until it hits the photocathode of the PM as a function of the angle of incidence. Obviously the ideal case of a two-dimensional cone with one reflection only is achieved within good approximation.

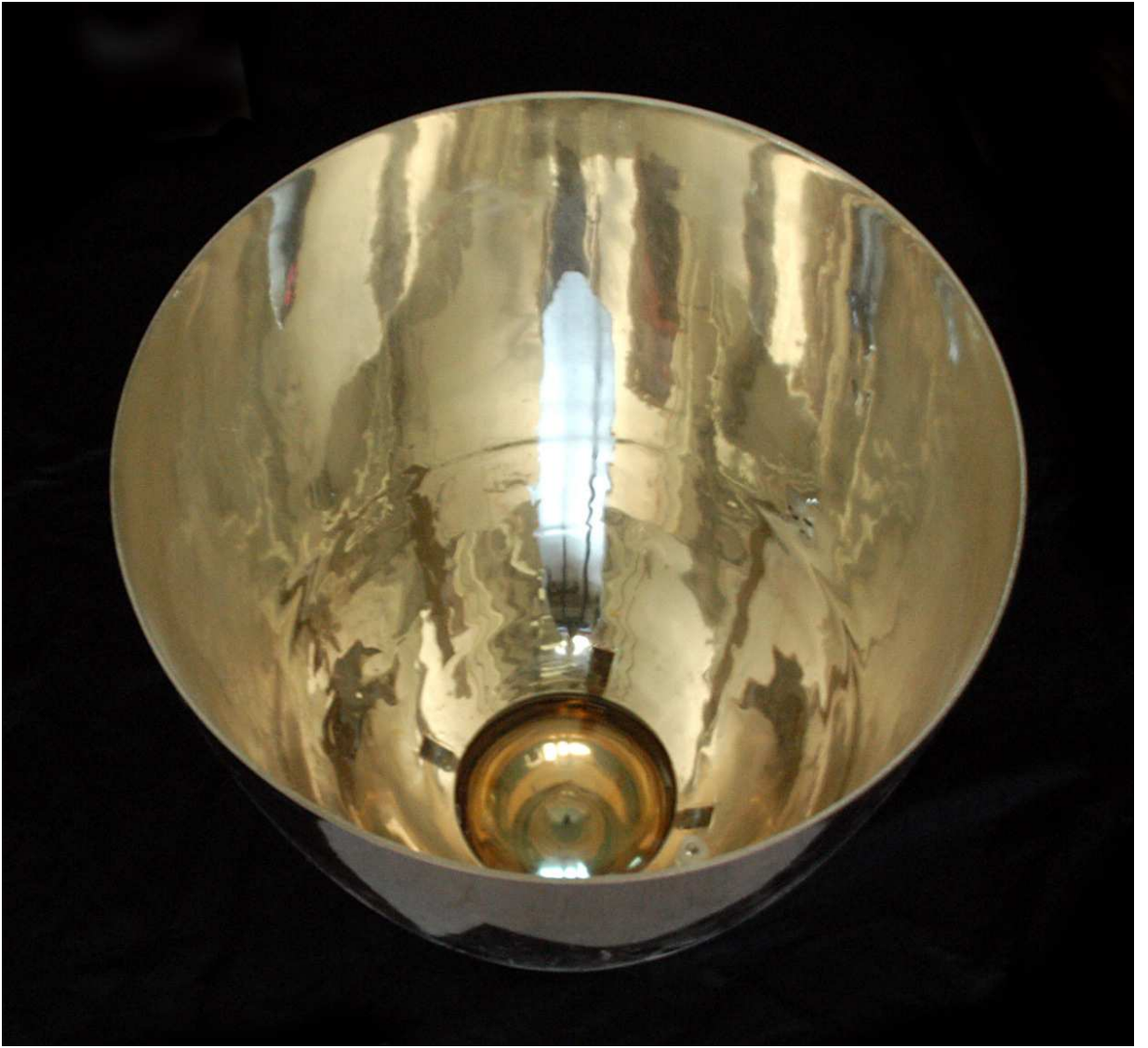

Fig. 5: CTF light concentrator mounted on PM.

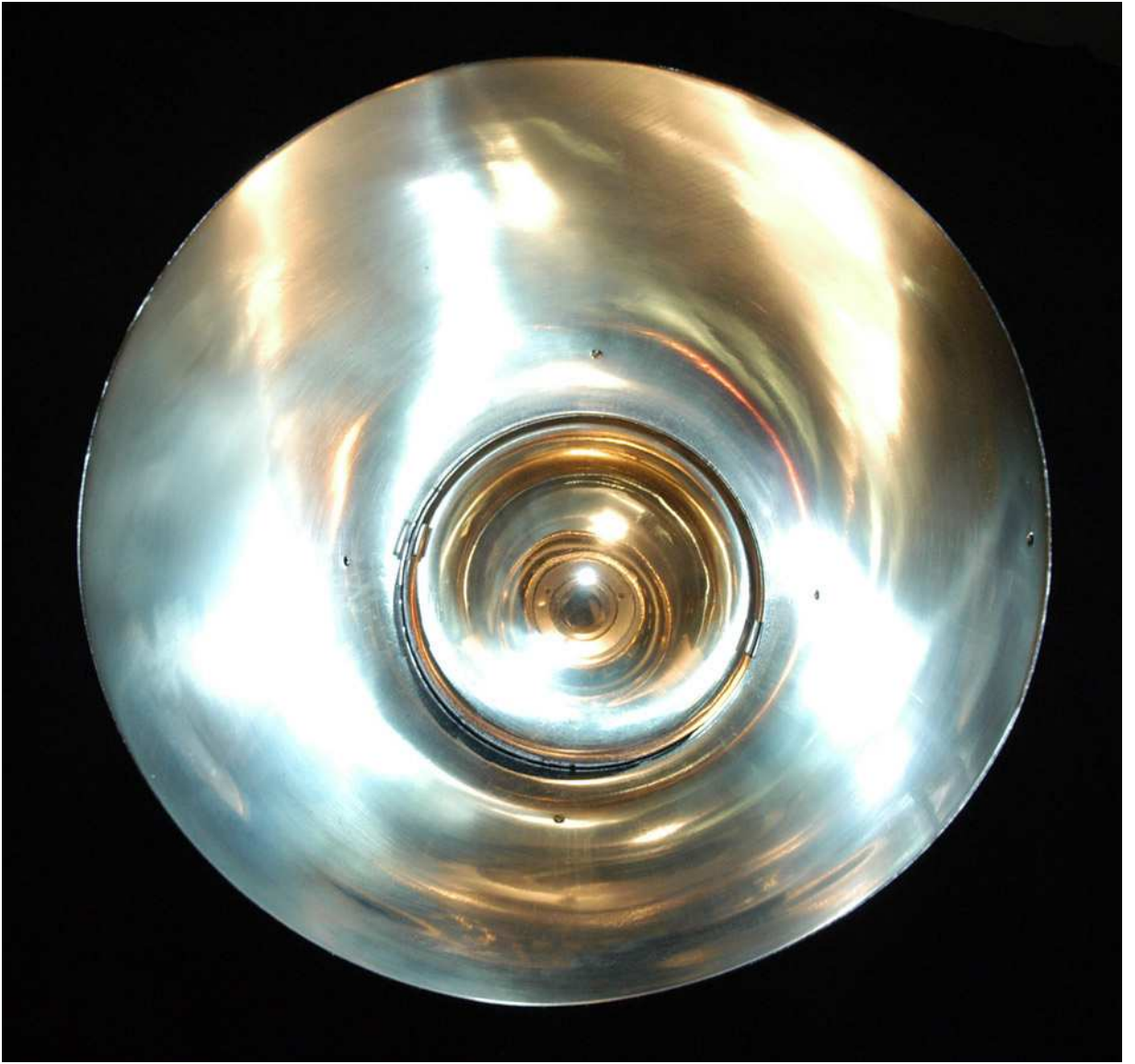

Fig. 6: Borexino light concentrator mounted on PM.